# Ejected Particles after Impact Splash on Mars: Aggregates and Aerodynamics

T. Becker[1], J. Teiser[1], T. Jardiel[2], M. Peiteado[2], O. Muñoz[3], J. Martikainen[3], J. C. Gomez Martin[3], J. Merrison[4], and G. Wurm[1]
[1] University of Duisburg-Essen, Faculty of Physics, Lotharstr. 1, 47057 Duisburg, Germany; timsven.becker@uni-due.de
[2] Instituto de Cerámica y Vidrio, CSIC, C/Kelsen 5, Campus Cantoblanco, 28049 Madrid, Spain
[3] Instituto de Astrofísica de Andalucía, CSIC, Glorieta de la Astronomía, s/n 18008 Granada, Spain
[4] Institute of Physics and Astronomy, Aarhus Universitet, 8000 Aarhus, Denmark
Received 2023 March 1; revised 2023 August 15; accepted 2023 August 21; published 2023 September 28

## Abstract

Our earlier laboratory measurements showed that low-velocity sand impacts release fine <5 $\mu$m dust from a Martian simulant soil. This dust will become airborne in the Martian atmosphere. Here, we extend this study by measuring aerodynamic properties of ejecta and characterizing deviations from the behavior of spherical, monolithic grains. We observe the settling of particles emitted as part of an impact splash. The sizes (20 to 280 $\mu$m) and sedimentation velocities (0.1 to 0.8 m s$^{-1}$) of the particles are deduced from high-speed videos while the particles sediment under low ambient pressure of about 1 mbar. The particles regularly settle slower than expected, down to a factor of about 0.3. Using optical microscopy, the shape of the captured particles is characterized by simple axis ratios (longest/smallest), which show that the vast majority of particles are irregular but typically not too elongated, with axis ratios below 2 on average. Electron microscopy further reveals that the particles are typically porous aggregates, which is the most likely reason for the reduction of the sedimentation velocity. Due to the reduced bulk density, aggregates up to 10 $\mu$m in diameter should regularly be a part of the dust in the Martian atmosphere.

*Unified Astronomy Thesaurus concepts:* Mars (1007); Planetary atmospheres (1244)

## 1. Introduction

From satellite and rover images, we know that there is a steady movement of soil occurring on many parts of the Martian surface (Metzger et al. 1999; Greeley et al. 2006; Silvestro et al. 2010; Chojnacki et al. 2011, 2019; Bridges et al. 2012a, 2012b; Reiss & Lorenz 2016), as well as dust movement in the Martian atmosphere (Cantor et al. 1999; Montabone et al. 2005). Even though it was long believed that the wind conditions on the Martian surface should only rarely be able to move soil (Pollack et al. 1976; Greeley & Iversen 1985; Sullivan et al. 2000; Haberle et al. 2003; Chojnacki et al. 2011), recent studies have showed that wind-induced shear stress might regularly be capable of moving sand-sized grains (Sullivan & Kok 2017; Andreotti et al. 2020). Wind tunnel experiments and theoretical studies based on wind tunnel data show that the particle size most susceptible to wind drag is about 100 $\mu$m or slightly larger under Martian conditions (Iversen et al. 1976; Greeley et al. 1980; Shao & Lu 2000; Swann et al. 2020). While for smaller grains, sticking (surface) forces are dominant (Greeley & Iversen 1985; Alfaro et al. 1997; Shao 2008; Kok et al. 2012; Rasmussen et al. 2015; Rondeau et al. 2015; Waza et al. 2023), larger grains are grounded by gravity. In the transition zone between these regions, particles are most susceptible to wind-induced shear stress. These 100 $\mu$m particles, which are set in motion at the minimum-threshold wind speed, however, are far too heavy to become entrained into the atmosphere (Kok et al. 2012). Instead, they start saltating, i.e., hopping along the surface (Shao 2008). At each reimpact upon the surface, they can eject smaller particles by breaking cohesive bonds (Bagnold 1941; Leach et al. 1989), enabling the wind to carry them along (Alfaro et al. 1997). So while micron-sized dust grains can hardly be liberated from the surface by wind-induced shear stress directly, impacts during saltation can liberate such dust particles. Particles of this size are thought to be suspended by turbulent eddies and carried farther upward within the atmosphere by means of convective flows (Edgett & Christensen 1991; Daerden et al. 2015; Haberle et al. 2017; Musiolik et al. 2018; Neary & Daerden 2018).

Our previous study analyzed this process for individual impacts—i.e., that bonds are broken down to the (sub-)micrometer scale, as we could detect particles even below 1 $\mu$m in diameter being ejected from slow saltating impacts (Becker et al. 2022). In that study, the final estimate of how much dust per saltating impact could go into long-term suspension in the atmosphere assumed a cutoff diameter of 3 $\mu$m as the upper limit. The cutoff diameter was chosen based on the results of studies using spectral data, performed by, e.g., Pollack et al. (1995), Tomasko et al. (1999), Wolff & Clancy (2003), Clancy et al. (2003), Lemmon et al. (2004), and Wolff et al. (2006). However, those studies only considered monolithical, spherical, or cylindrical grains for the estimation of the particle size. Natural particles are by no means perfectly spherical to begin with, though (Dietrich 1982; Ming & Morris 2017). Apart from monolithic particles not being spherical, aggregates being present is another factor (Sullivan et al. 2008; Waza et al. 2023). For these kinds of particles, this upper limit might not hold true.

If the particles were to be cluster–cluster aggregates (CCAs) with low fractal dimension, sedimentation speeds would be much lower than for nonfluffy, spherical particles of the same size (Meakin 1987; Nakamura et al. 1994; Wurm & Blum 2000). Thus, even large aggregates may enter into long-term suspension. Another possibility is aggregates being spherical, but with high porosity. Such porous material has already been identified on the Martian surface by, e.g., Moore & Jakosky (1989) and Sullivan et al. (2008). The effectively reduced density of a particle compared to a monolithic grain would increase the maximum size for airborne particles as well (Dietrich 1982), yet this would not be quite as extreme as for CCAs.

Morphology is not only important for the lifting process. It also plays a major role for grains entering into saltation. After all, if saltating particles are aggregates, it changes the nature of the reimpact into the soil (Shao 2008). Upon impact the aggregate can be damaged or even break down into a large number of fragments, that are small compared to the original aggregate (Kun & Herrmann 1999; Shao 2001, 2008; Kok et al. 2012) and can be carried away by the wind more easily. Kok (2011) even gives an analytical model, describing the particle size distribution (PSD) of particles generated by such fragmentation processes.

Furthermore, size and morphology also impact the retrieval of particle properties from remote sensing (Muñoz et al. 2021; Lin et al. 2023). Thus, having a clearer picture on what kinds of physical properties are expected for particles that are ejected by certain dust-generation mechanisms can help in further increasing the accuracy of the information gained from remote-sensing data. For example, a recent work by Martikainen et al. (2023) updates the optical properties of Martian dust by assuming more realistic particle shapes to simulate the Martian regolith, compared to the common assumption of particles being spheres or cylinders.

The question remains: what kinds of properties for nonspherical dust particles would be realistic for specific dust-generation mechanisms? Recently, for example, Waza et al. (2023) found large aggregates to be removed by wind on a thick dust layer. For the mechanism of impact-generated dust in a mix of large grains and small dust, we connect to our earlier studies on particle ejection during impacts and extend these experiments in this study. We now measure the sedimentation speed and determine the aerodynamic properties of the particles that would regularly be considered too large to become entrained. We connect these measurements to particles' aggregate structure and shape.

## 2. Experiments

In order to get insight into the structure and aerodynamics of the particles liberated in saltating impacts, we approached the problem from two sides. First, we measured the sedimentation velocity and the geometrical size for ejected particles. Second, we caught sedimenting particles and analyzed their shape under an optical microscope and used a scanning electron microscope (SEM) for high-resolution images. To obtain the aspired data, we used an extended version of the setup from our previous work (Becker et al. 2022).

### 2.1. Impact Setup

The experimental setup consists of a plunger that accelerates large saltators into a dust bed containing Martian soil simulant (see Figure 1(b)). The saltators have sizes between 180 and 250 $\mu$m and are accelerated to speeds of $1.04\,\mathrm{m\,s^{-1}} \pm 0.2\,\mathrm{m\,s^{-1}}$. These speeds are in agreement to wind shear velocities at the saltation threshold (Swann et al. 2020). Impact angles were also kept unchanged compared to our previous work, with $18.8° \pm 2.5°$, a value well within the range proposed by several studies (Bagnold 1941; Chepil 1945; White & Schulz 1977; Jensen & Sorensen 1986). The whole setup was placed within a vacuum chamber. The experiments were carried out at a pressure of 1 mbar. The pressure is at the lower end of the surface pressures on Mars, but was mostly chosen for technical reasons. For larger pressures, small particles settle too slowly and are easily picked up by convective flows. For smaller pressures, particles are mostly free falling in the laboratory and not reacting to the gas flow on reasonable timescales or distances.

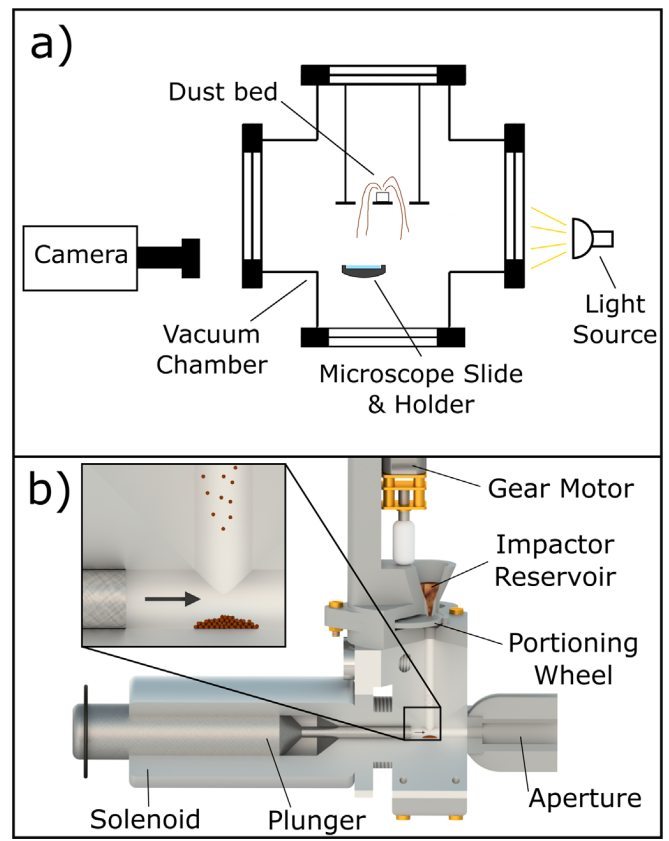

**Figure 1.** Sketch of the impact setup and acceleration unit (not to scale). (a) Experiment setup within the vacuum chamber in front view. The launcher is located behind the dust bed (not shown in the sketch). The brown lines schematically illustrate the sedimentation paths of the ejected particles. (b) Cross section of the launcher that launches saltators onto the dust bed.

To each side of the dust bed a recess was cut. Grains ejected during impacts fall through and sediment farther downward, as indicated in Figure 1(a). The particles were observed at 2000 fps using a NAC MEMRECAM HX-3 high-speed camera at a vertical distance of 3.6cm below the top of the dust bed. The field of view was $4544 \times 2556\ \mu$m, which means that falling particles could be observed over a maximum distance of 2.56mm. The resolution of the camera was 3.55 $\mu$m/px and the error was determined to be 2 pixels to each side. Farther down the vacuum chamber, a microscope slide was placed to capture the sedimenting particles. They were analyzed using an AxioCam ICc 1 mounted on a ZEISS Axioimager.M2 microscope, with a resolution of 0.37 $\mu$m/px and an error of 1 pixel to each side. A total number of ≈30 impact experiments were carried out and analyzed. Additionally, we used an SEM to gain information on the morphological structure of the individual sedimented particles.

### 2.2. Soil: Martian Simulant

For the experiment, we reused the Martian Global Simulant (MGS) sample (Cannon et al. 2019) we used in our previous work (Becker et al. 2022) as well. That is, we produced a bimodal size distribution of fine clay-sized particles and larger sand-sized saltators. Using this method, we tried to imitate the steady mixing of fine particles sedimenting from the Martian atmosphere and larger particles moved by wind through saltation. If we assume a regular mixing in that manner, the surface layer of the soil would be comprised of such a bimodal PSD. We set up a 1:1 mixture of small (<20 $\mu$m) particles and a larger fraction centered around roughly 100 $\mu$m. The conditioned sample was then placed in the dust bed. As impactors, particles of 180–250 $\mu$m from the MGS reservoir were used.

## 3. Methods

### 3.1. Sedimentation Speed and Aerodynamic Size

From the high-speed videos, particle tracks have been analyzed using the ImageJ plugin Trackmate (Ershov et al. 2022). Positions and timestamps were then used to fit a linear model over the trajectory and extract the sedimentation velocity from it. The videos also provided us with the time after launch at which each particle came into the field of view, as well as the particle diameters. The diameters were measured by hand for each particle individually at two different times to get an average.

From the measured particle sizes and sedimentation velocities, we determined the aerodynamic size, i.e., the size of a spherical grain of the same material density that corresponds to a given sedimentation speed $v_{Sed}$. The aerodynamical size can be determined from the particles' equation of motion. In the case of our experiments, the force $F_{Sed}$ that acts on a particle is

$$F_{Sed} = F_G - F_S, \tag{1}$$

where $F_G$ is gravity and $F_S$ is the Stokes drag. At high pressure, the Stokes drag is given as

$$F_S = 6\pi\mu R v_{Sed}(t), \tag{2}$$

with the particle radius $R$ and dynamic viscosity $\mu = 1.8 \cdot 10^{-5}$ Pa s for air. However, as we carry out the experiments at low pressure, a correction factor (Cunningham 1910) has to be applied, due to the transition from continuum to free molecular flow. This correction can be characterized by the Knudsen number $Kn = \frac{\lambda}{R_g}$, where $\lambda$ is the mean free path of the gas molecules and $R_g$ is the geometrical size. One parameterization of the correction factor is (Hutchins et al. 1995)

$$C = 1 + Kn \cdot (\alpha + \beta \cdot e^{-\frac{\omega}{Kn}}), \tag{3}$$

with $\alpha = 1.231 \pm 0.0022$, $\beta = 0.47 \pm 0.0037$, and $\omega = 1.178 \pm 0.0091$ as empirical constants. The corrected Stokes drag $F_{Sc}$ is then

$$F_{Sc} = \frac{F_S}{C}. \tag{4}$$

Using this and the gravitational force

$$F_G = mg = \rho_p \frac{4}{3}\pi R^3 g \tag{5}$$

with the material density $\rho_p$ then gives

$$v_{Sed}(R, t) = -\frac{1}{\gamma} \cdot (1 - e^{-\gamma t}), \tag{6}$$

with

$$\gamma = \frac{9\,\mu}{2\rho_p R^2 g C}. \tag{7}$$

We measured the average material density $\rho_p = 2460\,\text{kg/m}^3$ with a pycnometer and a scale.

Using the measured velocity and time from the camera images for each particle, we then numerically solved Equation (6) for $R$, which in this case is the corresponding aerodynamical radius $R_a$, and formed the ratio of aerodynamical to geometrical size for each particle.

### 3.2. Shape Distribution

From the optical microscope images, we retrieved data regarding the shape of the particles. We identified the longest and smallest axis in the cross sections of captured particles and formed the corresponding axis ratio. Additionally, we determined the diameter of a circular particle with the same cross section (mean diameter) for all 5318 individual particles.

## 4. Results

### 4.1. Sedimentation Speed and Aerodynamic Size

The data for the sedimentation velocity are shown in Figure 2. A total of 92 data points are presented, representing particles from 20 $\mu$m to 300 $\mu$m in diameter.

The dashed horizontal line shows the maximum velocity that a frictionless free-falling particle can reach, if it starts at rest at the height of the dust bed.

Besides the expected trend that larger particles sediment faster, Figure 2 shows a considerable spread of sedimentation velocities for similar particle sizes. The sedimentation velocities for same-sized particles can differ by a factor of up to 5 for small particles <50 $\mu$m. While this factor goes down for increasing particle sizes (e.g., to about 3 for particles around 100 $\mu$m), this deviation still implies that the geometrical size is not the only aspect determining the sedimentation velocity. To better evaluate the geometrical particle sizes and their corresponding sedimentation velocities, we determined the ratio of aerodynamical size and geometrical size. The ratio in dependence from the geometrical size is plotted in Figure 3. Here, a value smaller than 1 means that the particle's drag corresponds to a particle smaller than the observed geometrical particle size (slow regime). It can be seen that, apart from three exceptions, all data points are within the slow regime, reaching down to ratios below 0.3. For reference, the blue dotted line within the plot at a ratio of 0.71 corresponds to a spherical aggregate with a porosity of 50% (50% of the aggregate volume consists of solid material). The majority of particles being below that line implies that most observed particles are highly porous in nature. Figure 4 shows the distribution of porosities over the whole data set, apart from the three exceptions that have an aerodynamical-to-geometrical size ratio greater than 1 and consequently would have a porosity smaller than 0% as well, which obviously does not make sense. Those three particles are most likely ellipsoidal, monolithic grains, well aligned to the airflow.

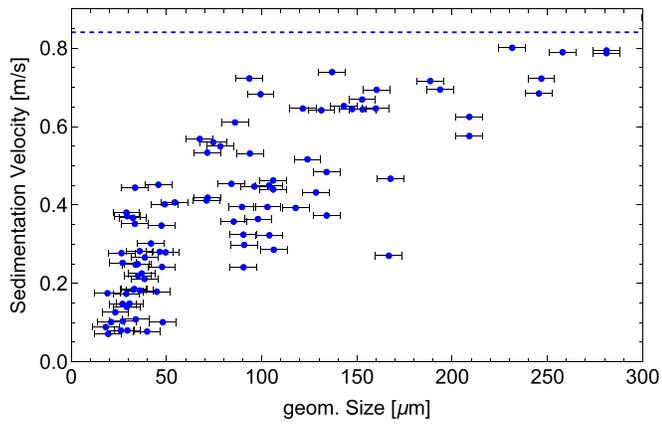

**Figure 2.** Sedimentation velocity over geometrical size (measured particle diameter) of the 92 observed particles. The dashed blue line is the freefall velocity from the dust bed height. For small geometrical sizes, the measured sedimentation velocity varies by up to a factor of 5 between nearly same-sized particles, while it decreases for increasing geometrical sizes.

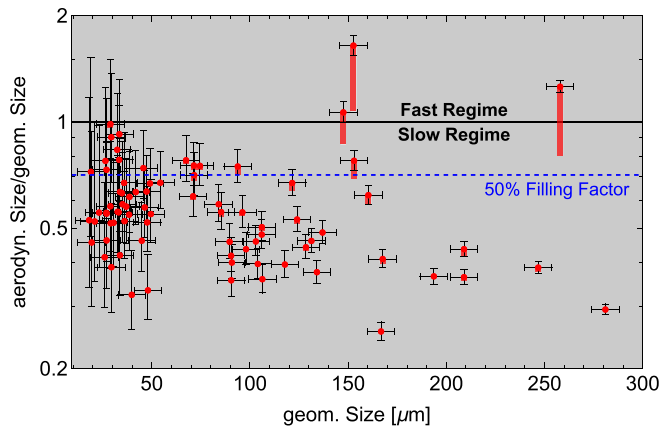

**Figure 3.** Ratio between aerodynamical and geometrical size. Particles with values smaller than 1 are slower than expected for their geometrical size. The dashed blue line represents aggregates with 50% porosity (50% of the aggregate volume is devoid of solid material). The black error bars represent the error caused by measurements of the geometrical size and the red error bars show the error for a + 5% shift in time.

### 4.2. Scanning Microscopy

This interpretation of highly porous aggregates coincides with SEM images of captured sedimented dust. Due to limited resources, we could only take eight images in total, of which three are presented in Figure 5. The full sample set is shown in the Appendix (see Figure 7). Based on our observations, we analyzed the morphological structure of the particles and categorized them into three different types: small aggregates, large aggregates shattered upon impact, and monolithic particles. An example of each type is shown in Figure 5.

A small aggregate of 20–30 $\mu$m in diameter can be seen in Figure 5(a). The visible part mainly consists of single particles on the order of 1 $\mu$m. Some small grains and aggregates of around 5 $\mu$m next to it suggest that they have been shed off upon impact on the sample holder.

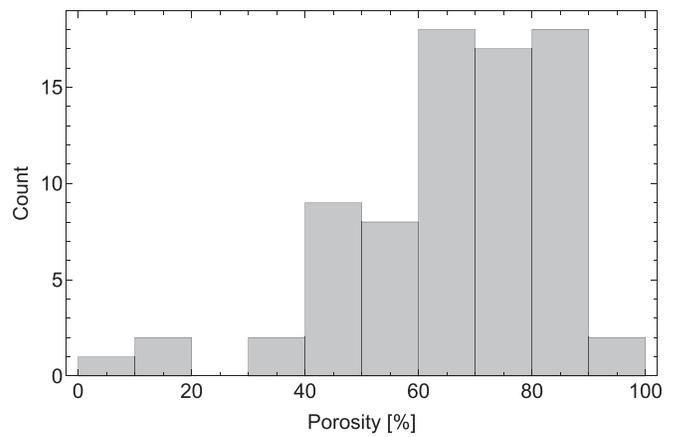

**Figure 4.** Histogram of the ratio of aerodynamical size to geometrical size for each particle mapped to the corresponding porosity.

Figure 5(b) shows a large area covered in small aggregates and monolithic particles of <5 $\mu$m up to >20 $\mu$m. The overall thin coverage of the sample holder compared to the high particle density and locality of this particular patch of material leads us to believe that these are the fragments of a large aggregate that shattered upon impact on the sample holder. We estimate the original aggregate to have been on the order of 100 $\mu$m.

Last, Figure 5(c) shows an example of a monolithic particle about 80 $\mu$m in size. Only a small number of micron-sized grains adhere to its surface. Additionally, it can be seen that the particle has a slightly prolonged shape.

These three cases represent the different types of particles observed. Having small and large aggregates as well as monolithic particles as ejecta matches the data presented in Figure 3.

### 4.3. Shape Distribution

All particles—even if they are highly porous—are overall rather round in shape. They are not perfectly spherical, though. Since a particle's shape also impacts the sedimentation speed, strongly elongated particles could either fall slightly faster or slower than a sphere with analogous average geometrical diameter, depending on their alignment to the gas flow. Cluster–cluster-aggregates (CCAs), for example, tend to align with the longest axis downward (Wurm & Blum 2000). If compact aggregates do so as well, they would fall slightly faster than their spherical counterparts, due to the cross section being subjected to the headwind being reduced. To investigate this topic, we plotted the axis ratio of the longest and smallest axes of each particle over the diameter of a circular particle with the same cross section in Figure 6.

Out of all data points, 96.3% lie below an axis ratio of 3 and 86.6% lie below an axis ratio of 2. Furthermore, there is a tendency that the y-axis spread of the data points decreases toward larger grain sizes, keeping in mind that the number of data points also decreases with increasing particle size. As the microscope footage does not discern between aggregates and monolithic particles, we cannot give specified information on the shape distribution for either population, but, in general, the particles are not very elongated. Thus, even if the alignment is not perfect, we only expect a small effect on the sedimentation speed due to irregular shapes.

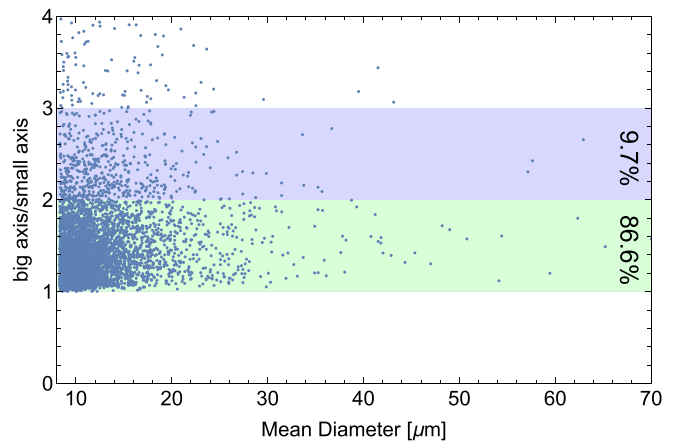

**Figure 6.** Ratios of the longest to smallest axes of particles from optical microscope images over surface equivalent diameter. The area between long axis-to-small axis ratios of 1 and 2 (highlighted in green) contains 86.6% of all particles, and the area between long axis-to-small axis ratios of 2 and 3 (highlighted in blue) contains 9.7% of all particles.

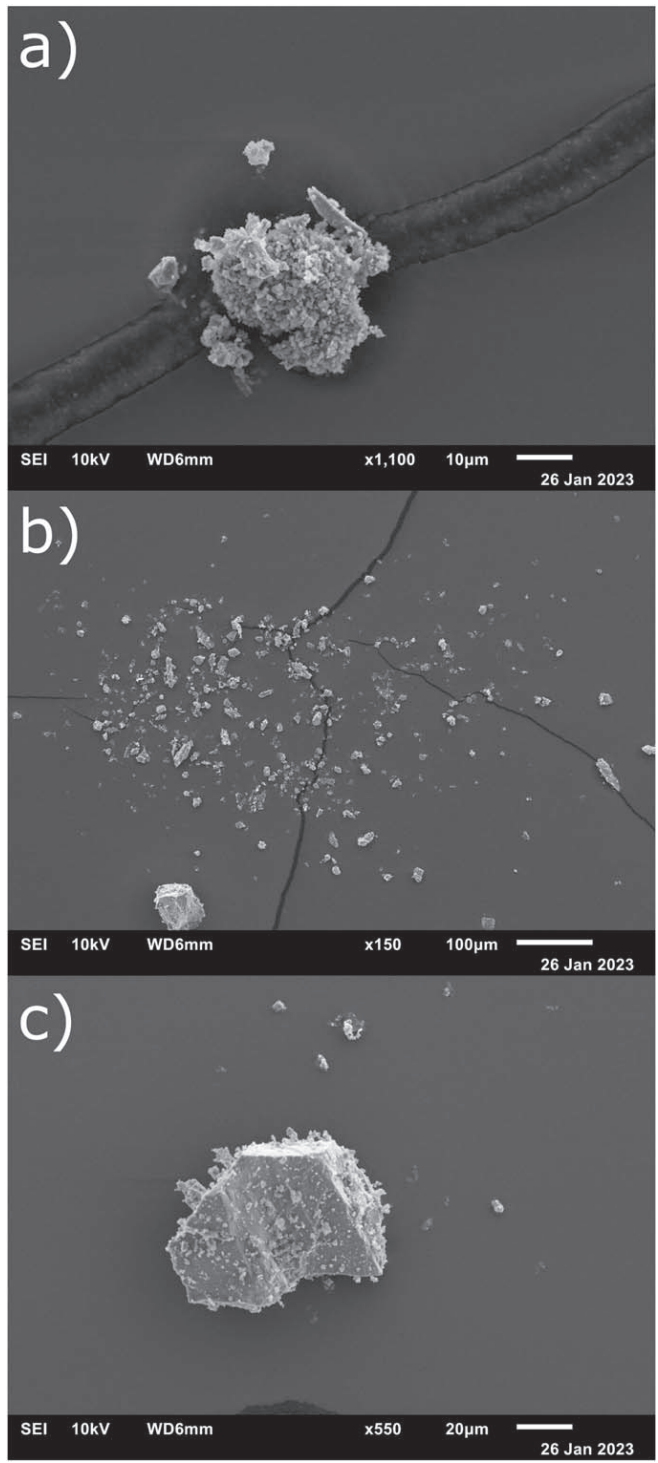

**Figure 5.** SEM images of captured particles. The black lines in the background are cracks in the adhesive coating. (a) A mostly intact aggregate of about 20–30 $\mu$m in diameter. Individual grains have sizes down to 1 $\mu$m and less. (b) An aggregate of about 100 $\mu$m that fragmented upon impact. (c) A monolithic particle of about 80 $\mu$m with a little micron-sized dust attached.

## 5. Caveats

We do not follow individual grains from ejection to impact. Therefore, we need to assume that particles start their fall from the same height at the top of the dust bed. The aerodynamical sizes of larger grains would need to be adjusted accordingly, as the measured times between lift and observation could be slightly larger. As indicated by the red error bars in Figure 3, which show the effect of a +5% shift in time on the measurements, such an adjustment would only considerably affect particles with a large geometrical size and a high ratio of aerodynamic to geometric size. Small particles reach terminal velocity quickly. Thus, a small change in timing does not affect the results noticeably. The same holds for very slow particles, such as particles with a large geometrical size, but a low ratio of aerodynamic to geometric size. Consequently, as can be seen in Figure 3, a small shift in time does not noticeably change the majority of our measurements and therefore does not change our conclusions either.

## 6. Conclusions

We analyzed sedimentation velocities for particles ejected from slow saltating impacts. We find a considerable spread of sedimentation velocities for similar particle sizes. The spread can be as large as a factor of 5 between the slowest and fastest sedimentation velocity for same-sized particles.

The shape analysis of the sedimented particles in Figure 6 shows that the overwhelming majority of particles are rather round in shape, having a long axis-to-small axis ratio of less than 2. Based on this finding, we would expect particles of the same geometrical size to have very similar sedimentation velocities, if they were monolithic. Thus, we see the discrepancy in the sedimentation velocities of same-sized particles in Figure 2 as an indication of particles being aggregates rather than having complex shapes. This argumentation is further supported by our findings from Figure 3. It shows that most particles have aerodynamical sizes that are smaller than their geometrical sizes, i.e., the particles sediment slower than they would be expected to. For some particles, the ratio of aerodynamical size to geometrical size even reaches down to values below 0.5, while for most particles it still is below 0.7. Keeping Figure 6 in mind, particle shape is a very unlikely cause of such low values. Taking a monolithic particle with an axis ratio of 2:1, the extreme values for its ratio of aerodynamical to geometric size would be 1.41 and 0.71, which are still greater than the ratios for most particles in Figure 3. Consequently, the findings from Figure 3 strongly point to particles being porous aggregates. The SEM images align with that reasoning as well, since they show aggregates as well as monolithic particles, with all of them generally having rather round shapes. Though the results from SEM data are only of a qualitative nature, they match the results from the shape and sedimentation speed analyses well. Thus, taking all our findings into account, we argue that the main reason for the discrepancy in the sedimentation velocities of the same-sized particles in Figure 2 is a difference in the porosity of the aggregates and monolithic particles, rather than an increase in the complexity of the shapes.

Ejecta from saltating impacts contain monolithic particles as well as aggregates, and they are not expected to have extreme shapes or morphologies. We know that for monolithic particles, 3 $\mu$m (effective diameter) is widely considered to be the upper limit to be suspended on Mars. As for the aggregates, we can estimate an upper size limit for long-term suspension using Figure 3. The lowest ratios of aerodynamical to geometric size are around 0.3. If we apply this ratio to a monolithic 3 $\mu$m grain to correspond it to an aggregate with the same aerodynamical size, we get an aggregate with a geometrical size of about 10 $\mu$m. Thus, the size limit for aggregates would be 10 $\mu$m, based on our estimation.

Applied to Mars or other planets with a sufficiently thick atmosphere, this shift in the geometric size of suspendable particles influences the total mass budgets of transported dust and optical properties. Since we do not know what fraction of the overall ejecta is constituted by aggregates, we cannot give a specific value for the impact on mass budgets. However, given our findings, an upward shift for the estimated mass is to be expected. Thus, saltation on Mars could bring a wider range of particle sizes into long-term suspension in the atmosphere than previously thought. The shift would have an effect on the optical properties of airborne dust as well, since the size and scattering behavior of aggregates vary from those of monolithic grains. What that means in detail would have to be investigated by further research, though. For future Mars missions, in situ measurements close to the Martian surface and in the atmosphere would be of great interest for us, to determine if mass budgets and circulation models have to be updated, as our study implies.

## Acknowledgments

This project has received funding from the European Union's Horizon 2020 research and innovation program under grant agreement No. 101004052. We thank the two anonymous referees for the thorough review of our manuscript.

## Appendix
## Supplemental Material

Here we add a Figure with the complete set of SEM pictures taken, sorted by case. The first column of Figure 7 represents Figure 5.

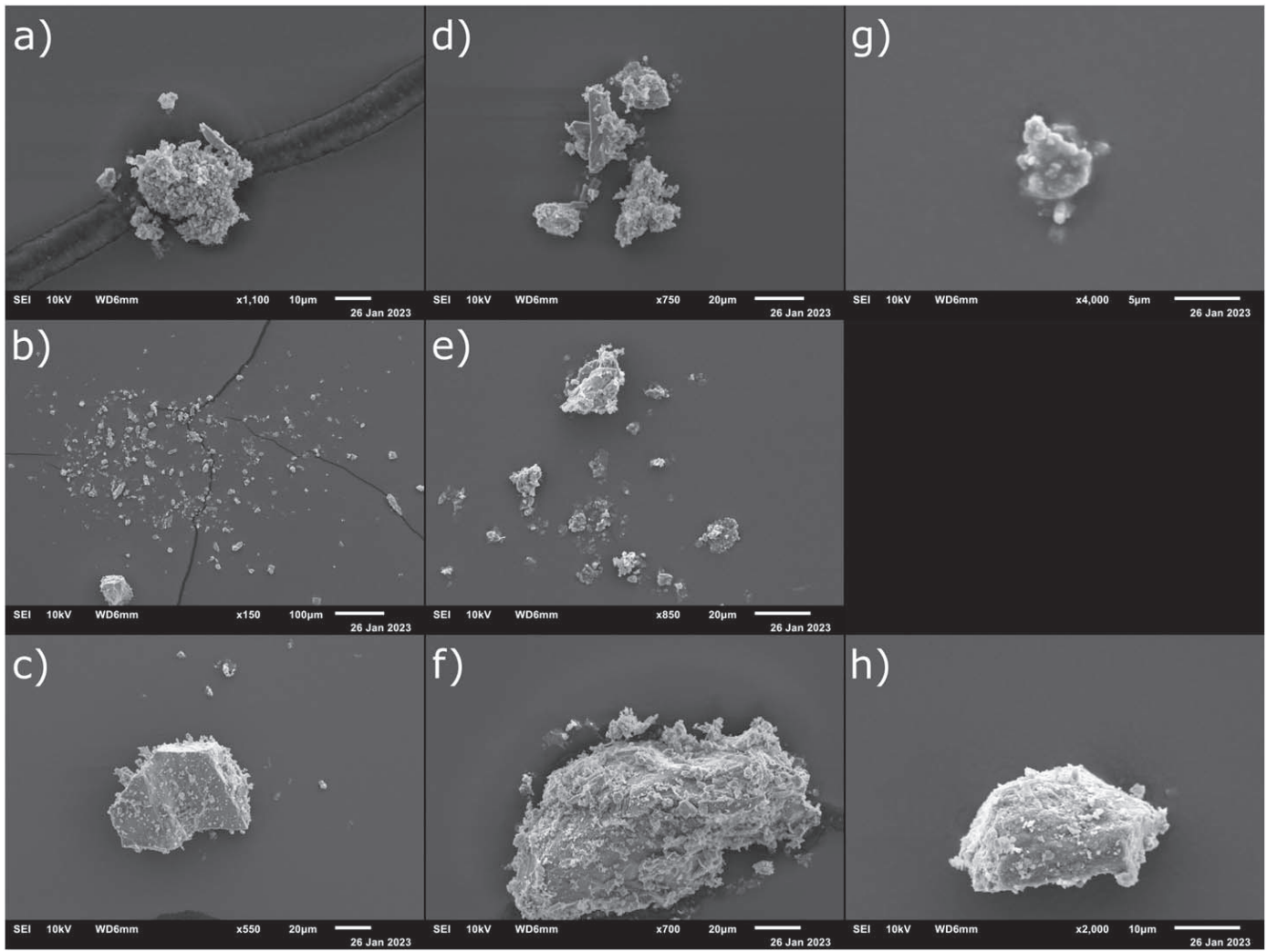

**Figure 7.** Compilation of all eight SEM images, with panels (a), (d), and (g) in the first row showing small aggregates, panels (b) and (e) in the second row showing what we think are aggregates shattered upon impact, and panels (c), (f), and (h) in the bottom row showing monolithic particles with clay-sized dust adhered to them.

**ORCID iDs**

O. Muñoz ● https://orcid.org/0000-0002-5138-3932
J. Martikainen ● https://orcid.org/0000-0003-2211-4001
G. Wurm ● https://orcid.org/0000-0002-7962-4961